\documentclass[traditabstract]{aa}
\usepackage{graphicx}
\usepackage{txfonts}
\usepackage{booktabs}
\usepackage{multirow}
\usepackage{natbib}
\usepackage{wasysym} 
\bibpunct{(}{)}{;}{a}{}{,}
\newcommand{\teff}{\mbox{$T_{\rm eff}$}}
\newcommand{\logg}{\mbox{$\log g$}}
\newcommand{\vsini}{\mbox{$v \sin i_{\tiny \star}$}}
\newcommand{\teq}{\mbox{$T_{\rm eq}$}}
\newcommand{\mictrb}{\mbox{$\xi_{\rm t}$}}
\newcommand{\mactrb}{\mbox{$v_{\rm mac}$}}

\newcommand{\kms}{\mbox{km\,s$^{-1}$}}
\newcommand{\halpha}{\mbox{$H_\alpha$}}

\newcommand{\rhostar}{\ensuremath{\rho_\star}}
\newcommand{\rhosun}{\ensuremath{\rho_\odot}}
\newcommand{\rhoj}{\ensuremath{\rho_{\rm J}}}
\newcommand{\rhopl}{\ensuremath{\rho_{\rm pl}}}
\newcommand{\rj}{R\ensuremath{_{\rm J}}}
\newcommand{\mj}{M\ensuremath{_{\rm J}}}
\newcommand{\me}{M\ensuremath{_\oplus}}
\newcommand{\rsun}{R\ensuremath{_\odot}}
\newcommand{\msun}{M\ensuremath{_\odot}}
\newcommand{\rpl}{\ensuremath{R_{\rm pl}}}
\newcommand{\mpl}{\ensuremath{M_{\rm pl}}}
\newcommand{\rstar}{\ensuremath{R_\star}}
\newcommand{\mstar}{\ensuremath{M_\star}}
\newcommand{\degree}{\mbox{\ensuremath{^\circ}}}   

\def\secos{$\sqrt{e} \cos \omega$}
\def\sesin{$\sqrt{e} \sin \omega$}
\def\feh{[Fe/H]}

\begin{document}

\title{Discovery of WASP-65b and WASP-75b: \\Two Hot Jupiters Without Highly Inflated Radii} 
\titlerunning{Discovery of WASP-65b and WASP-75b}
   
  \author{
 Y. G\'omez Maqueo Chew \inst{\ref{warwick},\ref{vandy},\ref{crya}} 
  \and F. Faedi \inst{\ref{warwick}}  
  \and D. Pollacco \inst{\ref{warwick}} 
  \and D. J. A. Brown \inst{\ref{sta}} 
  \and A. P.\ Doyle \inst{\ref{keele}} 
  \and A. Collier Cameron \inst{\ref{sta}} 
  \and M. Gillon \inst{\ref{liege}} 
  \and M. Lendl \inst{\ref{genv}}
  \and B. Smalley \inst{\ref{keele}} 
  \and A. H. M. J. Triaud \inst{\ref{genv},\ref{mit}} 
  \and R. G. West \inst{\ref{warwick}} 
  \and P. J. Wheatley \inst{\ref{warwick}}
  \and R. Busuttil \inst{\ref{open}}
  \and C. Liebig \inst{\ref{sta}}
  \and D. R. Anderson \inst{\ref{keele}} 
  \and D. J. Armstrong\inst{\ref{warwick}} 
  \and S. C. C. Barros \inst{\ref{lam}} 
  \and J. Bento  \inst{\ref{warwick},\ref{macq}} 
  \and J. Bochinski \inst{\ref{open}}
  \and V. Burwitz \inst{\ref{mpe}}
  \and L. Delrez \inst{\ref{liege}}
  \and B. Enoch \inst{\ref{sta}}
  \and A. Fumel \inst{\ref{liege}} 
  \and C. A. Haswell \inst{\ref{open}}
  \and G. H\'ebrard \inst{\ref{iap},\ref{ohp} } 
  \and C. Hellier \inst{\ref{keele}} 
  \and S. Holmes \inst{\ref{open}} 
  \and E. Jehin \inst{\ref{liege}}
  \and U. Kolb \inst{\ref{open}}
  \and P. F. L. Maxted \inst{\ref{keele}} 
  \and J. McCormac \inst{\ref{ing},\ref{warwick}}
  \and G. R. M. Miller \inst{\ref{sta}}  
  \and A. J. Norton \inst{\ref{open}} 
  \and F. Pepe \inst{\ref{genv}} 
  \and D. Queloz \inst{\ref{genv},\ref{cambr}}
  \and J. Rodr\'iguez \inst{\ref{oam}} 
  \and D. S\'egransan \inst{\ref{genv}}
  \and I. Skillen \inst{\ref{ing}}
  \and K. G. Stassun \inst{\ref{vandy},\ref{fisk}} 
  \and S. Udry \inst{\ref{genv}}
  \and C. Watson \inst{\ref{qub}} 
     }

\institute{
  Department of Physics, University of Warwick, Coventry CV4 7AL, UK; \email{y.gomez@warwick.ac.uk} \label{warwick} 
  \and Physics and Astronomy Department, Vanderbilt University, Nashville, Tennessee 37235, USA \label{vandy} 
  \and Centro de Radioastronom\'ia y Astrof\'isica, UNAM, Apartado Postal 3-72, 58089 Morelia, Michoac\'an, M\'exico  \label{crya}
  \and School of Physics and Astronomy,  University of St Andrews, St Andrews, Fife KY16 9SS, UK \label{sta} 
  \and Astrophysics Group, Keele University,  Staffordshire, ST5 5BG, UK \label{keele}
  \and Observatoire astronomique de l'Universit\'e de Gen\`eve, 51 ch.\ des Maillettes, 1290 Sauverny, Switzerland \label{genv} 
  \and Universit\'e de Li\`ege, All\'ee du 6 ao$\hat {\rm u}$t 17, Sart Tilman, Li\`ege 1, Belgium \label{liege}
  \and Department of Physics, and Kavli Institute for Astrophysics and Space Research, Massachusetts Institute of Technology, Cambridge, MA 02139, USA \label{mit}
  \and Aix Marseille Universit\'e, CNRS, LAM (Laboratoire d'Astrophysique de Marseille) UMR 7326, 13388, Marseille, France \label{lam}
  \and Department of Physics and Astronomy, Macquarie University, NSW 2109, Australia \label{macq}
  \and Institut d'Astrophysique de Paris, UMR7095 CNRS, Universit\'e Pierre \& Marie Curie, France \label{iap}
  \and Observatoire de Haute-Provence, CNRS/OAMP, 04870 St Michel l'Observatoire, France \label{ohp}
  \and Department of Physics and Astronomy, University of Leicester, Leicester, LE1 7RH, UK \label{leic}
  \and Department of Physical Sciences, The Open University, Milton Keynes, MK7 6AA, UK \label{open}
  \and Max Planck Institut f\"ur Extraterrestrische Physik, Giessenbachstrasse 1, 85748 Garching, Germany \label{mpe}
  \and Isaac Newton Group of Telescopes, Apartado de Correos 321, E-38700 Santa Cruz de Palma, Spain \label{ing}
  \and Department of Physics, University of Cambridge, J J Thomson Av, Cambridge, CB3 0HE, UK \label{cambr}
  \and Observatori Astron\`omic de Mallorca, Cam\'i de l'Observatori s/n 07144 Costitx, Mallorca, Spain \label{oam}
  \and Department of Physics, Fisk University, Nashville, Tennessee 37208, USA \label{fisk}
  \and Astrophysics Research Centre, Queen's University Belfast, University Road, Belfast BT7 1NN, UK \label{qub}
}

   \date{Received 18 July 2013; accepted 27 September 2013}

  \abstract{
We report the discovery of two transiting hot Jupiters, WASP-65b (\mpl\ = 1.55 $\pm$ 0.16 \mj; \rpl\ = 1.11 $\pm$ 0.06 \rj), 
and WASP-75b (\mpl\ = 1.07 $\pm$ 0.05 \mj; \rpl\ = 1.27 $\pm$ 0.05 \rj). 
They orbit their host star every $\sim$2.311, and $\sim$2.484 days, respectively. 
The planet host WASP-65 is a G6 star (\teff\ = 5600 K, \feh\ = $-$0.07 $\pm$ 0.07, age $\gtrsim$ 8 Gyr);
WASP-75 is an F9 star (\teff\ = 6100 K, \feh\ = 0.07 $\pm$ 0.09, age $\sim$ 3 Gyr).  
WASP-65b is one of the densest known exoplanets in the mass range 0.1 and 2.0 \mj\ (\rhopl\ = 1.13 $\pm$ 0.08 \rhoj), 
a mass range where a large fraction of planets are found to be inflated with respect to theoretical planet models. 
WASP-65b is one of only a handful of planets with masses of $\sim$1.5 \mj, a mass regime surprisingly
underrepresented among the currently known hot Jupiters.  
The radius of WASP-75b is slightly inflated ($\lesssim$10\%) as compared to theoretical planet models with no core,  
 and has a density similar to that of Saturn (\rhopl\ = 0.52 $\pm$ 0.06 \rhoj).
}

    \keywords{planetary systems -- stars: individual: (WASP-65, WASP-75) 
	-- techniques: radial velocity, photometry}

   \maketitle
%

\section{Introduction}

Since the discovery of the first extrasolar planet around a main-sequence star, 51 Peg \citep{Mayor1995},
our understanding of planetary systems has dramatically evolved.
Planetary science, which was previously based solely on our own Solar System, must be able   
to explain the observed diversity in physical properties and trends in the known exoplanet population \citep[e.g.,][]{Baraffe2010,Cameron2011}.
An exceptionally valuable subset of the known extrasolar planets are those that transit the disc of their host star.
To date, there are over 300 confirmed transiting exoplanets in the literature\footnote{See http://exoplanet.eu/}.   
Most of these discoveries have been the product of ground-based surveys, of which  
the Wide Angle Search for Planets \citep[WASP;][]{Pollacco2006} has been the most successful, 
along with the HATNet Project \citep{Bakos2004}, OGLE-III \citep{Udalski2002}, 
TrES \citep{Alonso2004}, XO Project \citep{McCullough2005}, and KELT \citep{Pepper2007}.  
The space missions CoRoT \citep{Baglin2006}, and Kepler \citep{Borucki2010} have also significantly increased the number
of discovered transiting planetary systems, and have been able to find much smaller planets than those that have
been discovered from the ground, as well as multi-planetary and circumbinary transiting systems. 

With knowledge of the physical properties of the stellar host, the transisting system's particular orbital geometry allows us to measure 
both the actual mass (i.e., \mpl\ instead of \mpl$\sin{i}$) {\it and} the radius of the transiting planet \citep[e.g.,][]{Charbonneau2000}. 
The wide range of observed planetary radii and, in particular, 
the large fraction of close-in Jupiter-mass planets with anomalously bloated radii 
\citep[e.g.,][]{Fortney2010,Leconte2010,Laughlin2011} 
challenge planetary structure models.
Transiting planets allow us to probe the planetary structure by inferring the bulk composition of the planet 
from its mean density.   
For example, among the most bloated planets, WASP-17b \citep{Anderson2010a}, 
and HAT-P-32b \citep{Hartman2011}  
have mean planet densities \rhopl\ 
of 0.06, 
and 0.11~\rhoj, respectively,  
that are not able to be reproduced with standard core-less planet models 
which predict the largest planets for a given mass \citep[e.g.,][]{Baraffe2008,Fortney2007}.  
Thus, planetary inflation mechanisms, such as  stellar irradiation \citep{Guillot1996}, atmospheric circulation
\citep[e.g.,][]{Showman2002,Guillot2006}, tidal effects \citep[e.g.,][]{Bodenheimer2000,Jackson2008}, 
enhanced atmospheric opacities \citep{Burrows2007}, and ohmic heating \citep{Batygin2010,Wu2012} have been proposed to explain
these anomalously large radii \citep[see also][]{Baraffe2010}. 
However, a single mechanism has not been able to explain the entire range of observed radii, and it is possible
that a combination of them come into play, with some being more effective than others in differing environments/conditions.  
Thus, it is paramount to expand the sample of well-characterized transiting planets in order to understand the physical and environmental factors
that determine the surprising diversity in planetary radii and orbits that have been thus far discovered. 

In this paper, we present 
two newly identified transiting planets in 
the WASP Survey: 1SWASP J085317.82+083122.8, 
hereafter \object{WASP-65}; and 1SWASP J224932.56-104031.8, 
hereafter \object{WASP-75}.
The WASP discovery photometry is described in \S\ref{swobs}. Section~\ref{rvobs}
describes the spectroscopic follow-up observations that are used to determine the
radial velocities of the planet hosts and the spectroscopically determined stellar parameters.  
The high-cadence, follow-up photometry, detailed in \S\ref{fuobs}, includes
data from four different facilities.
We derive the stellar physical properties in \S\ref{star} and \S\ref{mone}, and the planetary properties  
via the simultaneous modelling of the radial velocities and the light curves, 
and the use of theoretical isochrones 
in \S\ref{mcmc}.
Finally, in \S\ref{discussion}, we discuss the implications of these new discoveries
in the context of the known planetary population.   
 

\section{Observations}\label{obs}

WASP-65 and WASP-75 have been identified in several all-sky catalogues which
provide broad-band optical and infrared photometry, as well as proper motion information. 
Coordinates, broad-band magnitudes and proper motion of the stars are taken 
from the Fourth U.S Naval Observatory CCD Astrograph Catalog \citep[UCAC4;][]{Zacharias2012}, and are given in Table~\ref{table1}.

\subsection{WASP Observations}\label{swobs}

The WASP North and South telescopes are located in La Palma (ING -
Canaries Islands, Spain) and Sutherland (SAAO - South Africa), respectively.
Each telescope consists of 8 Canon 200mm f/1.8 focal lenses coupled to
e2v 2048$\times$2048 pixel CCDs, which yield a field of view of
$7.8\times7.8$ square degrees with a corresponding pixel scale of 13\farcs7 \citep{Pollacco2006}.

WASP-65 and WASP-75 ($V=11.90$ and $11.45$ mag, respectively) are located in an equatorial region of sky
that is monitored by both WASP instruments.
The WASP observations have an exposure time of 30 seconds, and a typical cadence of 8 min. 
All WASP data for the two newly discovered planets were processed with the 
custom-built reduction pipeline described in \citet{Pollacco2006}. 
The resulting light curves were analysed using our implementation of the Box Least-Squares and SysRem detrending algorithms (see \citealt{Cameron2006}; \citealt{Kovacs2002}; \citealt{Tamuz2005}) to search for signatures of planetary transits. 
Once the targets were identified 
as planet candidates a series of multi-season, multi-camera analyses were performed to strengthen the candidate detection. 
In addition, different de-trending algorithms \citep[e.g.,][]{Kovacs2005} were used on the single season and 
multi-season light curves to confirm the transit signal and the physical parameters of the planet candidate. 
These additional tests allow a more thorough analysis of the stellar and planetary parameters derived solely from the WASP data 
and publicly available catalogues (e.g., UCAC4) 
thus helping in the identification of the best candidates, as well as to reject possible spurious detections. 

\begin{figure}
\centering
\includegraphics[width=0.5\textwidth]{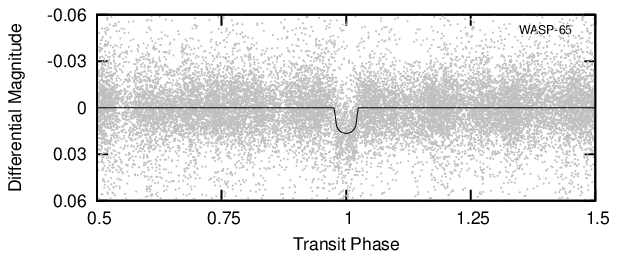}
\includegraphics[width=0.5\textwidth]{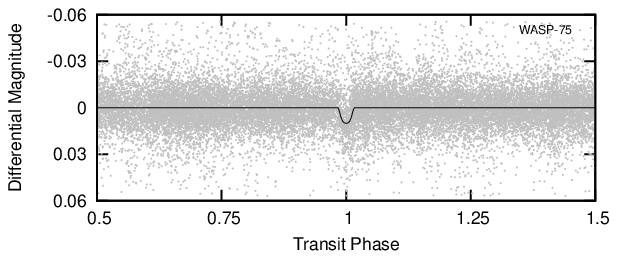}
 \caption{  Discovery WASP Light Curves. 
 {\em Upper panel}: WASP transit light curve of  WASP-65b, phase folded on the ephemeris given in Table~\ref{planet_params}. 
The black, solid line is the best-fit transit model, as described in \S\ref{mcmc}. 
{\em Lower panel}: Same as top panel for WASP-75b light curve.  
\label{swlcs}
}
\end{figure}

{\bf WASP-65} was first observed on 2009 January 14, and continued to be monitored over the   
following observing seasons up to 2011 April 24
with the WASP North facility.   
This resulted in a total of 18922 photometric points.  
The WASP data shows the periodic dip characteristic of a transiting planetary signal  
with a period of $P=2.31$~days, a transit duration of $T_{14}\sim2.6$~h, and a
transit depth of $\sim$11~mmag. 
The top panel of Fig.~\ref{swlcs} shows the discovery WASP photometry phase-folded over the
period derived in \S\ref{mcmc} superimposed on the model light curve. 

{\bf WASP-75} was observed with both the WASP telescopes from 2008 June 13
to 2010 November 11.  The WASP light curve is comprised of 23751 photometric data points.  
The WASP light curve is shown in Fig.~\ref{swlcs} (bottom panel) folded over
the identified transit period of $P=2.48$~days, and presents a transit depth of $\sim 10$~mmag, and
a transit duration of 1.91~h.

\begin{table}[t] 
\caption[]{Photometric and astrometric properties of the two stars WASP-65 and WASP-75}
\label{table1}
\begin{center}
\begin{tabular}{lccc}
\toprule \hline \\
 Parameter    & WASP-65 && WASP-75  \\
 \hline
 ${\rm RA (J2000)}$		& 08:53:17.83&& 22:49:32.57\\	
${\rm Dec (J2000)}$		& $+$08:31:22.8	&&$-$10:40:32.0\\	
${\rm B}$			&$12.57\pm0.01$	&&$12.05\pm0.03$ \\
${\rm V}$			&$11.90\pm0.04$	&&$11.45\pm0.01$\\
${\rm r}$			&$11.72\pm0.04$	&&$11.33\pm0.02$\\	
${\rm i}$			&$11.57\pm0.01$ &&$11.13\pm0.07$\\	
${\rm J}$			&$10.67\pm0.02$ &&$10.36\pm0.02$\\
${\rm H}$			&$10.44\pm0.03$ &&$10.10\pm0.02$\\	
${\rm K}$			&$10.35\pm0.02$ &&$10.06\pm0.03$\\	
 $\mu_{\alpha}$ (mas/yr)	&$3.8\pm1.3$	&&$42.8\pm1.9$	\\	
$\mu_{\delta}$   (mas/yr)	&$7.1\pm1.3$	&&$14.7\pm1.5$\\	
\bottomrule
\end{tabular}
\end{center}
{\footnotesize The broad-band magnitudes and proper motion are obtained from the UCAC4 catalogue.}
\end{table}
 
\subsection{Spectroscopic Follow-up Observations} \label{rvobs}

WASP-65 and WASP-75  
were observed with the CORALIE spectrograph mounted on the
1.2m Euler Swiss telescope at La Silla, Chile (\citealt{Baranne1996};
\citealt{Queloz2000}; \citealt{Pepe2002}). 
The data were processed with the CORALIE
standard data reduction pipeline. The radial velocity
uncertainties were 
derived from the photon noise. 
All spectra were single-lined.
For each planetary system the radial velocities of the host star were computed from a
weighted cross-correlation of each spectrum with a numerical mask of
spectral type G2, as described in \citet{Baranne1996} and
\citet{Pepe2002}. 
To test for possible false-positive scenarios, we performed
the cross-correlation with masks of different stellar spectral types
(i.e., G2 and K5) obtaining for each mask  similar radial
velocity variations. 

\begin{figure*}[ht]
   \centering
   \includegraphics[width=0.49\textwidth]{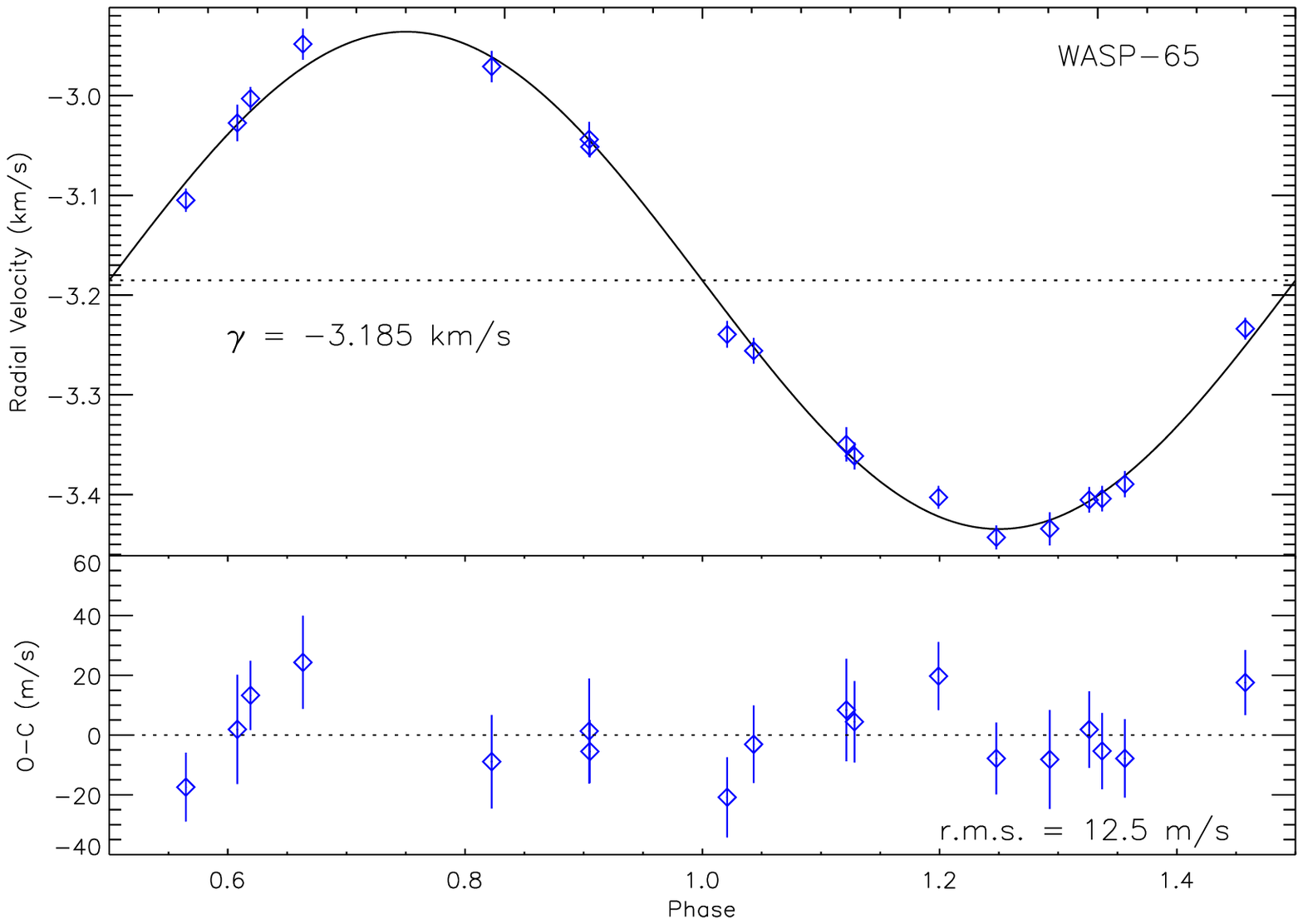}
   \includegraphics[width=0.49\textwidth]{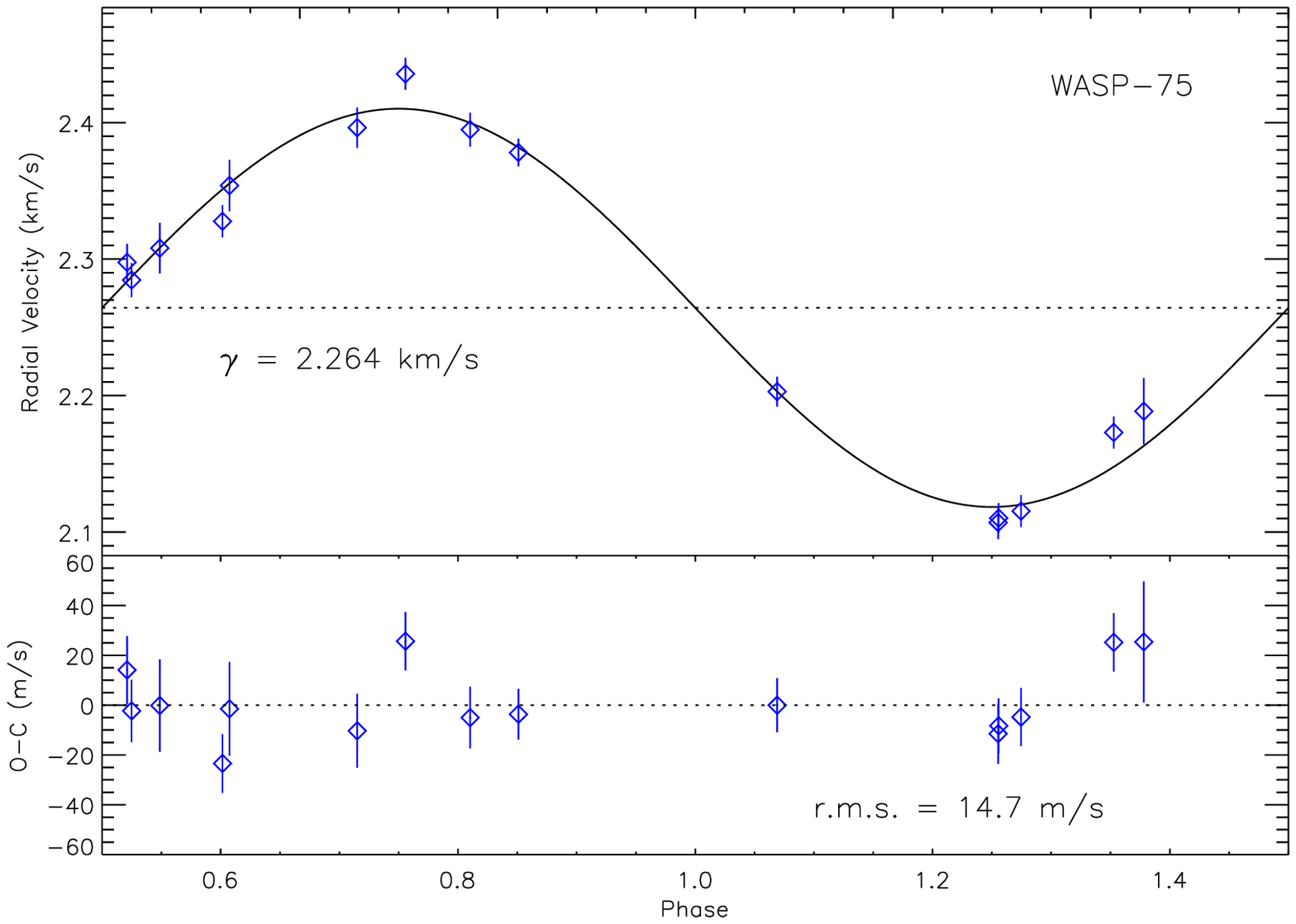}
      \caption{{\em Upper panels}: Phase folded radial velocity measurements of 
WASP-65 (left) and WASP-75 (right) obtained with the CORALIE
spectrograph. Superposed is the best-fit model RV curve with parameters from Table~\ref{planet_params}. 
The centre-of-mass velocity is marked by the horizontal, dotted line.  
{\em Lower panels}: 
Residuals from the radial velocity fit plotted against orbital phase; the dotted line in the lower panels marks zero. 
The residuals are in units of m\,s$^{-1}$.}
    \label{figrvs}%
   \end{figure*}

We present in Tables  \ref{W65_RVtable} and
\ref{W75_RVtable} the spectroscopic measurements of WASP-65 and WASP-75. 
Each table contains:  
the Barycentric Julian Date (BJD$_{\rm UTC}-$2\,450\,000.0), the stellar radial velocity (RV) measurements 
(\kms), the RV uncertainties (\kms), and the line bisector span measurements (V$_{span}$; \kms) as defined by \citet{Toner1988} 
and applied to the cross-correlation function as per \citet{Queloz2001}. 
The RV residuals (in units of m\,s$^{-1}$) to
the best-fit Keplerian model are found in the last column of Tables~\ref{W65_RVtable} and \ref{W75_RVtable};  
the residuals are calculated to have r.m.s. = 12.5~m\,s$^{-1}$ for
WASP-65, and r.m.s. = 14.7~m\,s$^{-1}$ for WASP-75.
Figure~\ref{figrvs}  shows the measured radial velocities and the residuals to the fit  
folded on the orbital period derived from the simultaneous analysis of the RVs and light curves (see \S\ref{mcmc}) for WASP-65 (left), and WASP-75 (right). 
Typical errors for the  CORALIE RV measurements are 10--15~m\,s$^{-1}$.


\begin{table} 
\caption{Radial Velocitiy Measurements of WASP-65  } 
\begin{center}
\begin{tabular}{ccccc}
\toprule \hline \\
BJD$_{\rm UTC}$ & RV & $\sigma_{\rm RV}$ & V$_{span}$& O -- C\\ 
$-$2\,450\,000 &(\kms)&(\kms)&(\kms)& (m\,s$^{-1}$) \\ 
\midrule
5683.53391	& -3.403	& 0.011	& -0.030& 20     \\ 
5696.53167	& -2.971	& 0.016	& -0.024& -9     \\ 
5706.48433	& -3.361	& 0.014	& -0.033& 4      \\ 
5711.48759	& -3.434	& 0.017	& 0.026 & -8     \\ 
5715.48188	& -3.239	& 0.013	& -0.046   & -21    \\
5721.46162	& -3.028	& 0.018	& 0.024	  & 2      \\
5722.46766	& -3.256	& 0.013	& -0.021    & -3     \\
5724.45871	& -3.044	& 0.018	& 0.005    & 1      \\ 
5725.45819	& -3.404	& 0.013	& -0.045   & -5     \\
5894.84370	& -3.003	& 0.012	& -0.050   & 13     \\
5917.83196	& -3.105	& 0.011	& -0.006   & -17    \\
5926.83061	& -3.234	& 0.011	& -0.028  & 18     \\
5927.86471	& -3.051	& 0.011	& -0.004    & -5     \\
5928.83795	& -3.405	& 0.013	& -0.029   & 2      \\
5958.70536	& -3.443	& 0.012	& 0.007    & -8     \\
6000.56135	& -3.389	& 0.013	& -0.036   & -8     \\
6003.58262	& -2.948	& 0.016	& 0.024  & 24     \\
6004.64137	& -3.350	& 0.017	& -0.035   & 8      \\
\bottomrule
\end{tabular}
\label{W65_RVtable}
\end{center}
{\footnotesize The columns are: the Barycentric Julian Date, the stellar RV measurements,
the RV uncertainties, the line bisector span measurements, and the residuals. }
\end{table}

\begin{table}[t] 
\caption{Radial Velocitiy Measurements of WASP-75}
\begin{center}
\begin{tabular}{ccccc} 
\toprule \hline \\
BJD$_{\rm UTC}$ & RV & $\sigma_{\rm RV}$ & V$_{span}$& O -- C\\ 
$-$2\,450\,000 &(\kms)&(\kms)&(\kms)& (m\,s$^{-1}$)\\ 
\midrule
5538.59311	& 2.396	& 0.015	& -0.009   & -10    \\
5795.80716	& 2.107	& 0.012	& 0.039     & -11    \\
5803.56482	& 2.189	& 0.024	& 0.005   & 25     \\ 
5804.63835	& 2.395	& 0.012	& 0.035    & -5     \\
5809.70786	& 2.378	& 0.010	& -0.043  & -4     \\
5810.76030	& 2.115	& 0.012	& 0.023    & -5     \\
5823.79344	& 2.298	& 0.014	& -0.026   & 14     \\
5830.58703	& 2.110	& 0.011	& 0.006     & -8     \\
5832.60740	& 2.203	& 0.011	& -0.003  & 0     \\
5833.73979	& 2.285	& 0.013	& 0.092   & -2     \\
5855.66998	& 2.173	& 0.012	& 0.029    & 25     \\
5856.67102	& 2.436	& 0.012	& 0.020   & 26     \\ 
5858.64072	& 2.308	& 0.018	& 0.044    & 0     \\
5883.62851	& 2.354	& 0.019	& -0.036   & -1     \\
5888.58231	& 2.328	& 0.012	& 0.018    & -23    \\ 
\bottomrule
\end{tabular}
\label{W75_RVtable}
\end{center}
{\footnotesize
The columns are: the Barycentric Julian Date, the stellar RV measurements,
the RV uncertainties, the line bisector span measurements, and the residuals. }
\end{table}

\begin{figure*}[ht]
   \centering
   \includegraphics[width=0.49\textwidth]{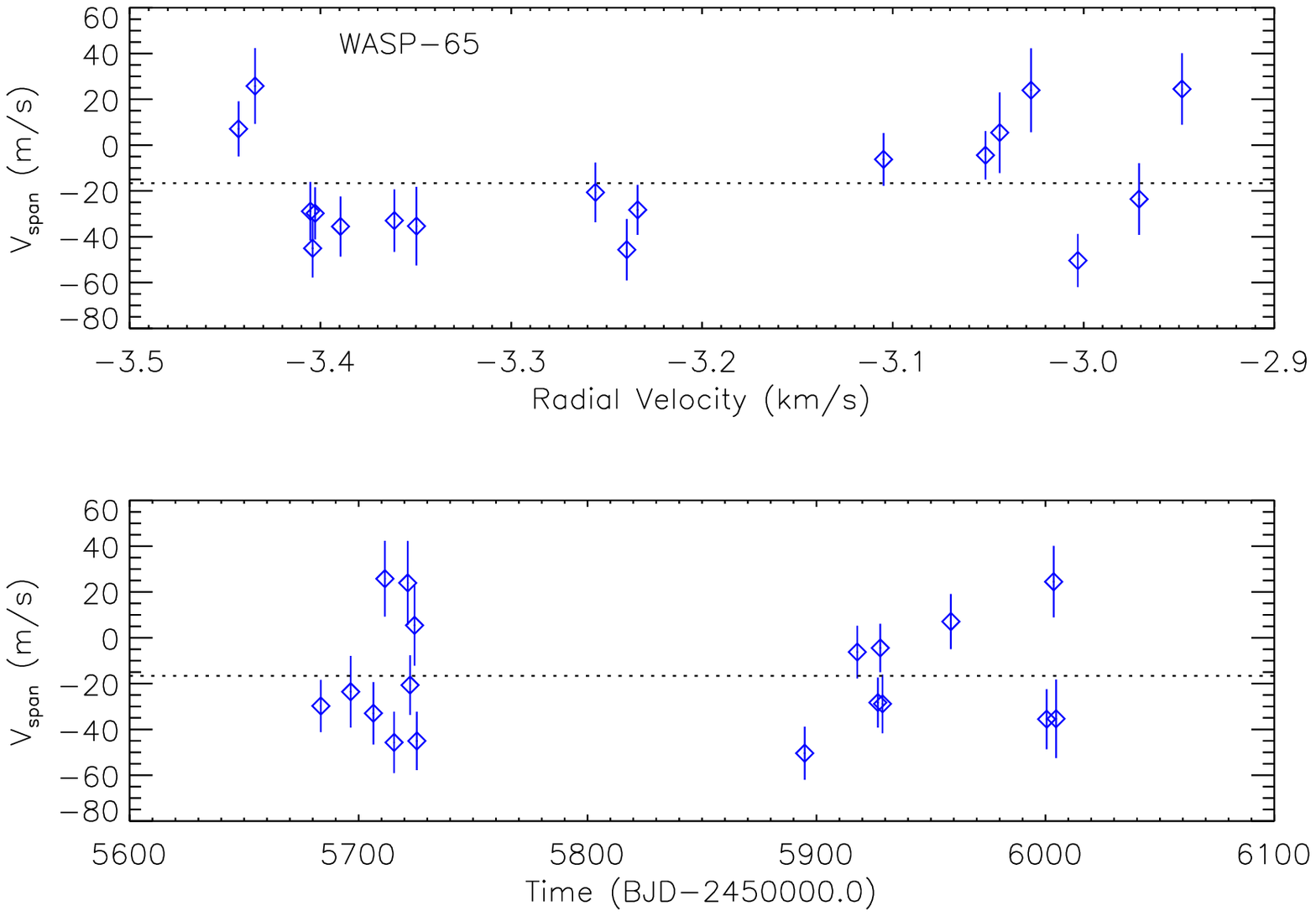}
   \includegraphics[width=0.49\textwidth]{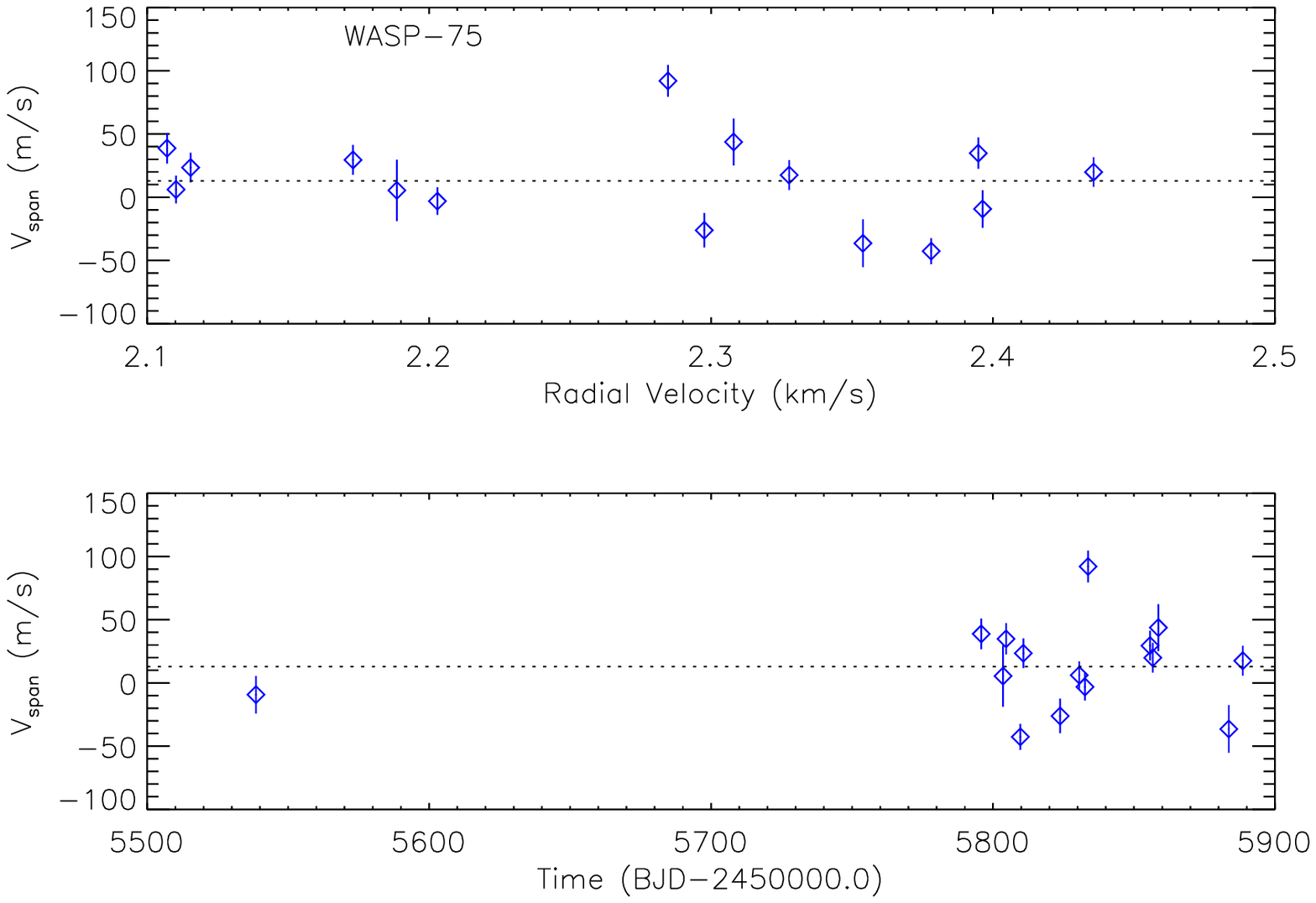}
      \caption{{\em Upper panels}: We show the bisector span measurements
 of WASP-65 (left) and WASP-75 (right) as a function of radial velocity. 
The horizontal, dotted line marks the mean line bisector span 
$<$$V_{span}$$> = -0.017$~\kms, and $<$$V_{span}$$> = 0.013$~\kms, respectively. 
{\em Lower panels}:  The bisector span measurements as a function of time 
(BJD$_{\rm UTC}$\,--\,2\,450\,000.0) for  WASP-65 (left) and WASP-75 (right). 
The bisector spans of both planet hosts are of the same order of magnitude
as the errors in the radial velocity measurements,
and show no significant variation nor correlation 
with radial velocity or time.   
This suggests that the radial velocity variations (with semi-amplitudes of $K_1$ = 0.249 $\pm$ 0.005 \kms\ for WASP-65b, and $K_1$ = 0.146 $\pm$ 0.004 \kms\ for WASP-75b) are due to Doppler shifts of the stellar lines induced by a planetary companion  
rather than stellar profile variations due to stellar activity or a blended eclipsing binary. 
}
    \label{figvspan}
   \end{figure*}

Furthermore, Figure~\ref{figvspan} presents the line bisector variations as a function of 
time  and of measured RV.  The line bisector measurements 
are examined to discard any false-positive scenarios that would reproduce the radial velocity motion 
of the star mimicking a planet signature that would induce a change in the line profile \citep[e.g.,][]{Dall2006}.  
Any asymmetries in spectral line
   profiles would be identified by the variation of the line bisector span, and could result from unresolved binarity or
   from stellar activity. Such effects would cause the bisector spans to
   vary in phase with radial velocity. No
   significant correlation is observed between either radial velocity and
   the line bisector (with a correlation coefficient of 0.04 for WASP-65, and -0.07 for WASP-75), or the bisector and the time at which
   observations were taken (with a correlation coefficient of 4e-06 for WASP-65, and 6e-05 for WASP-75). 
   This supports our conclusion that each signal originates from a 
   planetary companion as opposed to a blended eclipsing
   binary system, or to stellar activity \citep[e.g.,][]{Queloz2001}.

\begin{figure}[!ht]
   \centering
   \includegraphics[width=0.5\textwidth]{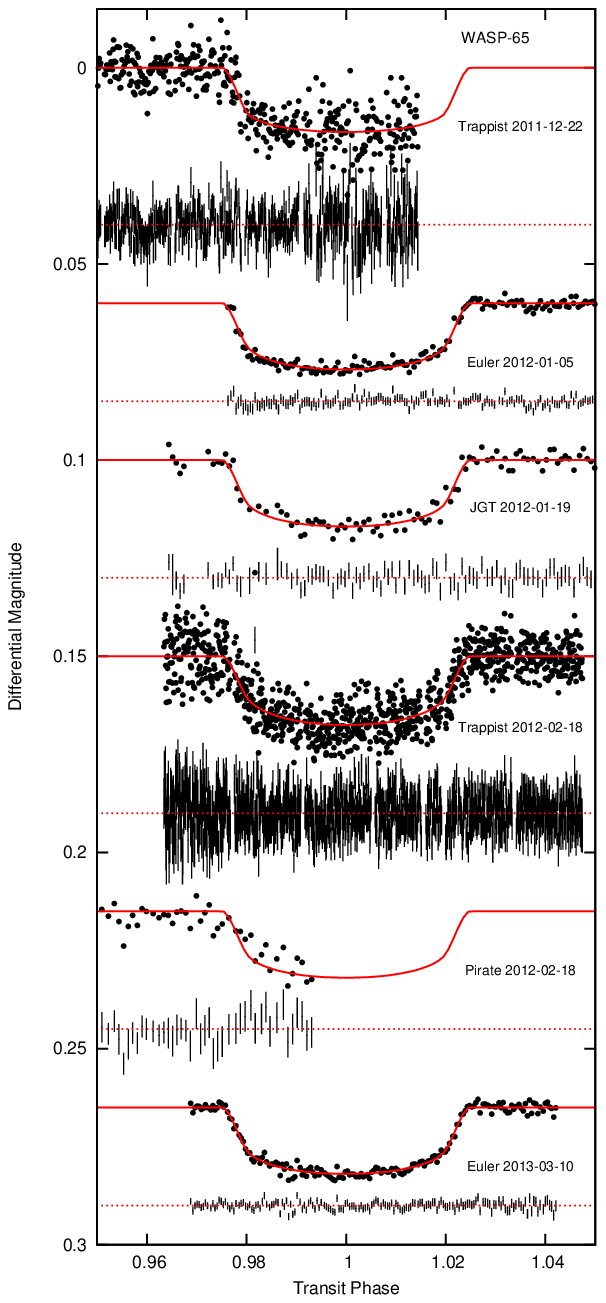} 
   \caption{ 
Follow-up, high-precision, time-series photometry of the WASP-65b during the transit (see Table~\ref{tablephot}). 
The observations are shown as red points and are phase folded on the ephemeris shown in Table~\ref{planet_params}.  
The superimposed, solid, black line is our best-fit transit model (\S\ref{mcmc}) using the formalism of \citet{Mandel2002}.
      The residuals from the fit and the individual data points photometric uncertainties 
are displayed directly under each light curve.
The light curves and residuals are displaced from zero for clarity. 
}
     \label{w65_lcs}
    \end{figure}

\begin{figure}[!ht]
   \centering
   \includegraphics[width=0.5\textwidth]{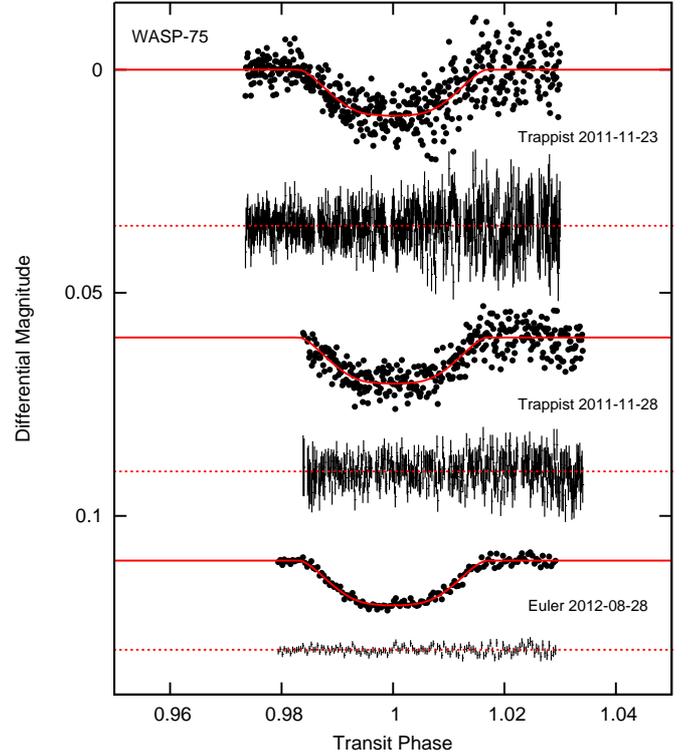}
   \caption{
Follow-up, high signal-to-noise light curves of WASP-75b during transit.  
Same as Fig.~\ref{w65_lcs}.  }
     \label{w75_lcs}
    \end{figure}

\subsection{Follow-Up Multi-band Photometry}\label{fuobs}

In order to better constrain the systems' parameters, 
high-cadence, high-precision time series photometry during the transits of WASP-65b and WASP-75b were obtained.
These follow-up light curves include data from four different telescopes (see below for details), and are summarized in Table~\ref{tablephot}.
All photometric data presented here are available electronically from CDS\footnote{Tables 8--11 are only available in electronic form at the CDS via anonymous ftp to cdsarc.u-strasbg.fr (130.79.128.5) or via http://cdsweb.u-strasbg.fr/cgi-bin/qcat?J/A+A/}. 
We show in Figures~\ref{w65_lcs} and \ref{w75_lcs}
the follow-up photometry for the transits of  
WASP-65b, and WASP-75b, respectively. In each plot we show the
differential magnitude versus orbital phase, along with the residual
to the best-fit model (see \S\ref{mcmc}). The data are phase folded on the ephemeris
derived by our analysis of each individual object as given in 
Table~\ref{planet_params}. 

\begin{table}[ht]
\caption{Follow-Up Time-Series Transit Photometry} 
\begin{center}
\begin{tabular}{cccc} 
\toprule \hline\\
Planet & Date & Instrument & Filter \\ 
\midrule
{\multirow{5}{*}{WASP-65b}}	
                                & 2011 12 22 & TRAPPIST & $I +z$ \\ 
				& 2012 01 05 & EulerCam & Gunn $r$ \\ 
                                & 2012 01 19 & JGT & $R_C$ \\ 
                                & 2012 02 18 & TRAPPIST &  blue blocking \\ 
				& 2012 02 18 & PIRATE & $R$ \\ 
				& 2013 03 10 & EulerCam & Gunn $r$ \\
\midrule
{\multirow{3}{*}{WASP-75b}}
				& 2011 11 23 & TRAPPIST & $I +z$ \\ 
                                & 2011 11 28 & TRAPPIST & $I +z$ \\ 
                                & 2012 08 28 & EulerCam & Gunn $r$ \\ 
\bottomrule
\end{tabular}
\label{tablephot}
\end{center}
\end{table}

\subsubsection{TRAPPIST Observations}

Both WASP systems were observed with TRAPPIST 
(TRAnsiting Planets and PlanetesImals Small Telescope; \citealt{Gillon2011})
located at ESO La Silla, Chile.   
Its thermoelectrically-cooled camera is equipped with a 2K$\times$2K
Fairchild 3041 CCD with a 22\arcmin$\times$22\arcmin~field of view (i.e., 0.65\arcsec/pixel). 
The observations were done using a readout mode of $2\times2$
MHz with $1\times1$ binning, resulting in a readout + overhead time of
6.1~s and a readout noise of 13.5 e$^-$. 
A slight defocus was
applied to the telescope to optimize the observation efficiency and to
minimize pixel-to-pixel effects. 
The TRAPPIST facility is described in detail by \citet{Jehin2011} and \citet{Gillon2011}. 
A standard pre-reduction
(bias, dark, flatfield correction), was carried out and the stellar
fluxes were extracted from the images using the IRAF/DAOPHOT aperture
photometry software \citep{Stetson1987}. After a careful selection of
reference stars, differential photometry was then obtained.

The transit of WASP-65b was observed twice with TRAPPIST.  One partial transit was observed on 2011 December 22 in the `I + z' filter 
(that has a transmittance $>90\%$ from 750~nm to beyond 1100 nm) 
with the observations starting about two hours before the transit's ingress, and ending at sunrise after mid-transit but before the start of the egress.   
As WASP-65 changed position in the night sky, 
the telescope had to undergo a ``meridian flip" which caused the field of view to rotate and the stars to change pixel position.   
We have accounted for this in our light curve analysis by treating the data before the ``meridian flip" as independent from the data after the ``meridian flip"
to allow for an offset.
The exposure time of these data was 20 s per frame with an average FWHM of 4 pixels for the stellar sources. 
A full transit of WASP-65b was observed on 2012 February 18 with the blue-blocking filter (with a transmittance >90\% above 500nm).  
The observations began more than 0.5 h before the ingress of the transit and ended over 1 h 
after egress.
The exposure time per frame was 8.0 s with an average FWHM of 4.4 pixels. 
There was also a ``meridian flip" during the observations of this transit.  

The WASP-75 system was also observed with TRAPPIST during the transit of the planet using
the {\it I+z} filter. 
The first TRAPPIST light curve was acquired on 2011 November 23 spanning the full transit, 
including $\sim$0.5 h before the transit ingress and $\sim$1~h after egress.  However,
the last half hour of data was obtained with an airmass of more than 1.8. 
The exposure time per frame was 15 s, with an average FWHM of 4.5 pixels.  
WASP-75 was observed again with TRAPPIST on 2011 November 28, obtaining photometry 
of an almost full transit.  The observations began shortly after ingress started and ended 
about 1~h after egress. 
The exposure time per frame was 20 s, with an average FWHM of 4.4 pixels.

\subsubsection{EulerCam Observations}

We observed two transits of WASP-65b and one transit of WASP-75b using an \textit{r'-Gunn} filter with EulerCam 
which is mounted on the 1.2~m Euler-Swiss telescope at the La Silla Observatory
in Chile.
All images were corrected for
bias and flat-field variations, and the light curves were obtained using differential aperture photometry. 
\citet{Lendl2012} describe in detail the EulerCam instrument, and the reduction procedures used.

The first observations of WASP-65b captured an almost complete transit (starting shortly after the beginning of the ingress) on 2012 January 06 (UT),
using fixed exposure times of 60~s. 
 The second set of observations took place on 2013 March 11 (UT) covering a complete transit, and fixed exposure times of 70s were used.
In both instances, the detector was read out through four ports in order to improve our efficiency,
and the telescope was slightly defocussed to improve the PSF sampling. 

A full transit of WASP-75b was observed with EulerCam with the \textit{r'-Gunn} filter on 2012 August 28 (UT). 
The observations started about half an hour before the beginning of ingress, and ended about an hour after egress.
A similar observing strategy was used as for the WASP-65b transits with the exposure time fixed at 60~s, with 
defocussing, and  
the detector readout through four ports.

\subsubsection{JGT Observations}

We observed the transit of WASP-65b across its host star with 
the 1-m James Gregory telescope (JGT) at the University of St Andrews Observatory in Scotland, UK.
More details of the telescope and instrument can be found in, e.g.,  
\citet{CollierCameron2010} and \citet{Hebrard2012}.  
A full transit was observed on 2012 January 19. 
The JGT light curve is comprised of 99 photometric measurements made using an $R_C$ (Cousins) filter with an exposure time of 150s each. 
The pre-eclipse out-of-transit measurements and part of 
ingress were affected by clouds, but the full width of transit, including times of first and last contact, were detected.
The stellar fluxes were extracted from the flat-fielded CCD frames using the photometry routines in the Starlink PHOTOM package. 
The differential photometry was calculated using a single nearby comparison star, which was the only object brighter than the target in the 15' field of view.

\subsubsection{PIRATE Observations}

The transit of WASP-65b was also observed on 2012 February 18 by the PIRATE Facility\footnote{\url{http://pirate.open.ac.uk/index.html}}  
(PIRATE Mk II configuration) 
located at the Observatori Astron{\`o}mic de Mallorca 
\citep[for details see][]{Holmes2011}.  
The time series is composed of 120-sec exposures that captured nearly 6 
hours of out-of-transit light curve (with the pier flip after about 2 hr), as well as the ingress. 
The conditions were good, giving a typical FWHM of 3.1\arcsec.
Additional out-of-transit and in-transit data were obtained with non-optimal weather conditions on 2012 January 19, 2012 March 02, and 2012 March 03
and were only used to constrain the ephemeris of the transit.  
All frames were taken with the Baader R filter; 
 this has a performance similar to the Astrodon Sloan $r'$ 
filter 
 used by the APASS survey \citep{Smith2010}. 
The data sets were calibrated in the standard way using flat field, dark and bias frames. We 
constructed the light curves using the ensemble photometry pipeline described in \citet{Holmes2011}. 
Pre- and post-pier flip branches were analysed separately.

\section{Results}

\subsection{Spectroscopically-determined Stellar Properties}\label{star}
For both planets the same stellar spectral analysis has been
performed by co-adding individual CORALIE spectra. 
The standard pipeline reduction products were
used in the analysis. The analysis was performed using the methods
given in \citet{Gillon2009}. The \halpha\ line was used to determine
the effective temperature (\teff). The surface gravity (\logg) was
determined from the Ca~{\sc i} lines at 6122{\AA}, 6162{\AA} and
6439{\AA} \citep{Bruntt2010a}, along with the Na~{\sc i} D and Mg~{\sc i} b lines. 
The parameters for WASP-65 and WASP-75 obtained from the spectral analysis are listed in
Table~\ref{Stellar_params}.

The elemental abundances were determined from equivalent width
measurements of several clean and unblended lines. A value for
microturbulence (\mictrb) was determined from Fe~{\sc i} using the
method of \citet{Magain1984}. The quoted errors are estimated to  
include the uncertainties in \teff, \logg\ and \mictrb, as well as
the scatter due to measurement (dependent on data quality), and atomic data uncertainties. 
The projected stellar rotation velocity (\vsini) was determined by fitting
the profiles of several unblended Fe~{\sc i} lines. For each system, 
the macroturbulence (\mactrb) was assumed based on the
calibration by \citet{Bruntt2010b}.   The telluric lines
around 6300\AA\ were used to determine the instrumental FWHM. 
There are no emission peaks evident in the
Ca~{\sc ii} H+K lines in the spectra of the two planet hosts. 
The parameters obtained for each planet host from the spectroscopic analysis are discussed below:

{\bf WASP-65:} 
A total of 10 individual CORALIE spectra of WASP-65 were co-added to produce a single spectrum
with a typical S/N of around 60:1. 
Our spectral analysis yields  
$\teff=5600 \pm 100$~K, $\logg=4.25 \pm 0.10$ (cgs), and \feh $=-0.07
\pm 0.07$ dex, and a spectral type of G6V. 
Taking into account the instrumental line profile (FWHM = 0.11
$\pm$ 0.01~{\AA}) and the macroturbulence (\mactrb\ = 2.0 $\pm$ 0.3 {\kms}),  
the projected stellar rotational velocity was determined to be \vsini\ = 3.6 $\pm$ 0.5 ~\kms.  
There is no significant detection of lithium in the spectra, with an equivalent
width upper limit of 12~m\AA, corresponding to an abundance upper limit of $\log
A$(Li) $<$ 1.14 $\pm$ 0.10. This implies an age of at least several Gyr \citep{Sestito2005}.
The rotation rate ($P = 17.9 \pm 3.4$~d) implied by the {\vsini} gives a
gyrochronological age of $\sim 1.72^{+1.26}_{-0.76}$~Gyr using the empirical relationship of 
\citet{Barnes2007}. 
The latter of these age indicators suggests that WASP-65 is a younger version of our Sun.  
However, assuming a higher value for \mactrb, like that from the calibrations of \citet{Valenti2005} or \citet{Gray2008}, 
\vsini\ would be lower and a longer rotation period would be derived implying an older age for the planet host.  
A better measure of the rotation period of WASP-65 would be from photometric rotational modulation; none was observable in the WASP light curve,
 which was searched with a sine-wave fitting algorithm, described in \citet{Maxted2011}. 
Moreover, given the lack of stellar activity (i.e., the absence of Ca~{\sc ii} H+K emission) and 
the comparison of the stellar properties with theoretical evolutionary models (see Table~\ref{tracks}), 
 WASP-65 seems to be older than our Sun ($\gtrsim$ 8 Gyr).

{\bf WASP-75:} 
A total of 15 individual CORALIE spectra of WASP-75 were co-added to produce a
single spectrum with a typical S/N of around 100:1. 
The derived spectroscopic properties of WASP-75 are $\teff=6100 \pm 100$~K, $\logg=4.5 \pm 0.1$ (cgs), and \feh $=0.07
\pm 0.09$ dex, and a spectral type of F9V. 
Considering a macroturbulence (\mactrb\ = 3.5 $\pm$ 0.3 {\kms}) and an instrumental FWHM of 0.11 $\pm$ 0.01~{\AA}, 
the best fitting value of \vsini\ = 4.3 $\pm$ 0.8 ~\kms\ was obtained.
The lithium line strength $\log A$(Li) = 2.52 $\pm$ 0.09  implies an age of approximately
2$\sim$5 Gyr according to \citet{Sestito2005}.
The rotation rate ($P = 11.7 \pm 2.7$~d) implied by the {\vsini} gives a
gyrochronological age of $\sim 1.69^{+1.58}_{-0.87}$~Gyr using the
\citet{Barnes2007} relation.
Both the gyrochronological and the age derived from the Li abundance 
imply that WASP-75 is a relatively young system, which is consistent with the age derived from
stellar evolutionary tracks ($\sim$3--4 Gyr; see Table~\ref{tracks}).

\begin{table}[t]
\caption{Stellar Properties of WASP-65, and WASP-75 from Spectroscopic Analysis}
\begin{center}
\begin{tabular}{lccc} 
\toprule \hline
Parameter		&WASP-65		&WASP-75 \\ 
\midrule
\teff~(K)&   5600 $\pm$ 100 K &   6100 $\pm$ 100 K \\	
\logg	&   4.25 $\pm$ 0.1 &   4.5 $\pm$ 0.1 \\
\mictrb~(\kms)	 &   0.9 $\pm$ 0.1 &   1.3 $\pm$ 0.1 \\
\vsini~(\kms)	&   3.6 $\pm$ 0.5 &   4.3 $\pm$ 0.8  \\
{[Fe/H]}		 &$-$0.07 $\pm$ 0.07 &   0.07 $\pm$ 0.09 \\
{[Na/H]}	&   0.08 $\pm$ 0.13 &   0.14 $\pm$ 0.05 \\
{[Mg/H]}	&   0.07 $\pm$ 0.08&   0.17 $\pm$ 0.15 \\
{[Si/H]}	 &   0.23 $\pm$ 0.06&   0.10 $\pm$ 0.09 \\
{[Ca/H]}	&   0.10 $\pm$ 0.15  &   0.11 $\pm$ 0.09 \\
{[Sc/H]}	 &   0.07 $\pm$ 0.10 &   0.19 $\pm$ 0.08 \\
{[Ti/H]}	 &   0.04 $\pm$ 0.07 &   0.10 $\pm$ 0.13 \\
{[V/H]}    &   0.05 $\pm$ 0.11 &   0.13 $\pm$ 0.09 \\
{[Cr/H]}	&   0.04 $\pm$ 0.13 &   0.10 $\pm$ 0.10 \\
{[Co/H]}	 &   0.14 $\pm$ 0.10&   0.15 $\pm$ 0.10 \\
{[Ni/H]}	&   0.05 $\pm$ 0.06&   0.08 $\pm$ 0.10 \\
$\log A$(Li)	 &   $<$ 1.14 $\pm$ 0.10  &   2.52 $\pm$ 0.09 \\
Sp. Type   	 &   G6  &   F9 \\
Distance~(pc)   &   310 $\pm$ 50 &   260 $\pm$ 70 \\ 
\bottomrule
\end{tabular}
\end{center}
\label{Stellar_params}
{\small {\bf Note:} 
Spectral Type estimated from \teff\
using the table in \cite{Gray2008}. }
\end{table}

\subsection{Stellar Mass Determination}\label{mone}

\begin{table}[ht!] 
\caption{Stellar Masses and Ages for WASP-65 and WASP-75} 
\begin{center}
\begin{tabular}{ccccc} 
\toprule \hline 
 & \multicolumn{2}{c}{WASP-65} & \multicolumn{2}{c}{WASP-75} \\
\midrule  \medskip
 & \multicolumn{4}{c}{\it Theoretical Evolutionary Models} \\
 & \mstar~(\msun) & Age (Gyr) & \mstar~(\msun) & Age (Gyr) \\
 \cline{2-5}\\ 
Padova$^1$ & 0.89$^{+0.16}_{-0.02}$&8.9$^{+3.7}_{-2.3}$&1.14$^{+0.04}_{-0.04}$&3.01$^{+1.33}_{-1.08}$ \smallskip\\
YY$^2$ & 0.93$^{+0.06}_{-0.06}$&8.8$^{+3.2}_{-2.9}$&1.17$^{+0.04}_{-0.03}$&3.12$^{+0.78}_{-0.95}$ \smallskip\\
Teramo$^3$ & 0.95$^{+0.05}_{-0.18}$&12.2$^{+3.5}_{-3.2}$&1.12$^{+0.07}_{-0.04}$&4.32$^{+1.63}_{-1.49}$ \smallskip \\
VRSS$^4$ & 0.90$^{+0.07}_{-0.04}$&11.2$^{+4.5}_{-3.8}$&1.13$^{+0.07}_{-0.05}$&3.45$^{+1.66}_{-0.84}$ \smallskip \\
\midrule \medskip 
 & \multicolumn{4}{c}{\it Empirical Relationship}\\ 
 & \multicolumn{2}{c}{\mstar~(\msun)}  & \multicolumn{2}{c}{\mstar~(\msun)}  \\
\cline{2-5} \\ 
Enoch$^5$ & \multicolumn{2}{c}{0.99 $\pm$ 0.02} & \multicolumn{2}{c}{1.15 $\pm$ 0.03} \\ 
\midrule \medskip 
 & \multicolumn{4}{c}{\it Adopted Stellar Mass}\\
 & \multicolumn{2}{c}{\mstar~(\msun)}  & \multicolumn{2}{c}{\mstar~(\msun)}  \\
\cline{2-5} \\ 
Mean & \multicolumn{2}{c}{0.93$^{+0.12}_{-0.16}$} & \multicolumn{2}{c}{1.14$^{+0.07}_{-0.06}$}\smallskip	\\
\bottomrule
\end{tabular}
\label{tracks}
\end{center}
{\footnotesize References. (1)~\citet{Marigo2008,Girardi2010}; (2)~\citet{Demarque2004}; (3)~\citet{Pietrinferni2004}; 
(4)~\citet{VandenBerg2006}; (5)~\citet{Enoch2010}. } 
\end{table} 

The absolute properties of the planet are well determined to the extent the 
stellar physical properties are accurate and precise.   
Thus, we have determined the stellar mass using four theoretical evolutionary models
and a stellar empirical calibration \citep{Enoch2010,Torres2010}, as described below.  The mean of these 
five independent mass estimates
is adopted for the rest of our analysis (see \S\ref{mcmc}), and its uncertainty is given
by the possible range of masses including the individual uncertainties.   
Table~\ref{tracks} summarizes the results from the interpolation of the four theoretical models, 
from the empirical relationship, and, lastly, the mean stellar mass adopted to derive the final orbital, stellar,
 and planetary properties.

   The stellar mass is derived from the spectroscopically-determined stellar 
effective temperature and metallicity (\S\ref{star}), and  
    the mean stellar density \rhostar, directly determined from transit light curves.  
   Transiting planets allow us to measure \rhostar\ independently from the \teff\ determined from the
   spectrum 
   (assuming $\mpl \ll \mstar$; see also \citealt{Seager2003}), 
as well as of theoretical stellar models
 \citep[e.g.,]{Sozzetti2007,Hebb2009}.  
We measured the mean stellar density of both planet hosts via  
our Markov-Chain Monte Carlo (MCMC) analysis   
   (see \S\ref{mcmc}).  
In the case of the theoretical evolutionary models, the four sets of tracks  
   used are: a) the Padova stellar models by \citet{Girardi2000}, b) the
   Yonsei-Yale (YY) models by \citet{Demarque2004}, c) the Teramo models
   of  \citet{Pietrinferni2004}, and d) the Victoria-Regina stellar
   models (VRSS) of \citet{VandenBerg2006}. 
The interpolation of the isochrones and mass tracks for the metallicity derived from the spectral analysis 
is done using a Delaunay triangulation \citep{Delaunay1934}, as
implemented by \citet{Bernal1988} and developped by \citet{Pal2006}. 
The errors are derived using the error ellipse from \teff\ and \rhostar$^{-1/3}$, 
taking into account the range of values given by the 1$-\sigma$ uncertainties, and the points at 45 degrees between these on the error ellipse.
Additionally, we have incorporated the uncertainty in the stellar mass due to the 1$-\sigma$ error in \feh.  
The stellar mass is also derived from the empirical calibration of \citet{Enoch2010} adapted
for transiting planets with measurable \rhostar\ from the study by  \citet{Torres2010} of  
 eclipsing binary stars with masses and radii known to better than 3\%. 
The uncertainty on the stellar mass derived from the empirical relationship results from our MCMC analysis (\S\ref{mcmc}). 
The stellar masses derived for WASP-75 
   from the four sets of stellar evolution models
   (Table~\ref{tracks}) agree very well with each other, and with the stellar mass from the Enoch calibration from our MCMC analysis, within 
their 1--$\sigma$ uncertainties.   
In the case of WASP-65, the \mstar\ from the empirical Enoch relationship is consistent 
within 2--$\sigma$ with the masses derived from the theoretical stellar tracks. 
For a more robust measurement of the stellar mass, we have calculated the mean 
of the stellar masses derived from the four theoretical models and the empirical Enoch calibration,
which is used as a prior in our MCMC analysis below and is used in the determination of the planetary properties. 
The uncertainty in the adopted stellar masses are given by the 1$-\sigma$ range for the individual derivations  
in order to account for all sources of error discussed above.

\subsection{Planetary Physical Properties}\label{mcmc}

\begin{table*}[!ht] 
\caption{System parameters of WASP-65 and WASP-75} 
\label{planet_params}
\begin{center} 
\begin{tabular}{lccc} 
\toprule \hline \\
&WASP-65b&WASP-75b&\\
\hline 
\\
$P$ 			 	& $  2.3114243  \pm 0.0000015$ 	& $2.484193  \pm  0.000003$ &d \\  
$T_{0}~^{\dagger}$ 			&$6110.68772   \pm  0.00015$ 	& $6016.2669    \pm  0.0003$ &d \\	
$T_{\rm 14}~^{\ddagger}$ \smallskip& $  0.11396 \pm  0.00045$ 	& $0.0822 \pm  0.0011$ &d \\
$T_{\rm 12}=T_{\rm 34}$ \smallskip & $0.0118^{+ 0.0004}_{-0.0003}$ 	& $0.030^{+ 0.003}_{-0.002}$  &d \\ 
$\Delta F=\rpl^{2} / \rstar^{2}$ 	& $  0.01280 \pm  0.00015$ 	& $0.0107 \pm  0.0003$ \\ 
$b$ \medskip &		$0.149  ^{+  0.082}_{-  0.095}$ &$ 0.882   ^{+  0.006}_{-  0.008}$& \rstar	\\ 
$i$ \medskip  		& $  88.8   ^{+  0.8}_{-  0.7}$ & $  82.0  ^{+  0.3}_{-  0.2}$ &$^\circ$ \\ 
$K_{\rm 1}$			& $0.249 \pm  0.005$ 		&  $0.146 \pm 0.004$ &\kms\\ 	
$\gamma$			& $-3.1853  \pm   0.0009$ 	& $  2.26429  \pm  0.00006$ &\kms \\ 
$e$ 								& 0.                                                                      & 0. & $fixed$\\                                                
\mstar 				& $  0.93    \pm  0.14 $ 	&$  1.14   \pm  0.07 $ &\msun \\	
\rstar 				& $  1.01   \pm  0.05 $  	& $  1.26   \pm  0.04 $ &\rsun \\
$\log g_{\star}$ 		& $4.40 \pm  0.02$ 		& $4.29 +  0.02 $ &cgs \\ 
\rhostar& $0.91 ^{+ 0.03}_{-0.04} $ 	& $0.56 \pm  0.04 $ &\rhosun \\		
\mpl				& $1.55  \pm 0.16$ 		& $1.07 \pm 0.05$ &\mj  \\ 	
\rpl \smallskip			&$ 1.112  \pm  0.059 $ 		& $ 1.270   \pm  0.048 $&\rj \\
$\log g_{\rm pl}$ \smallskip 	& $ 3.458  ^{+  0.014 }_{-  0.018 }$ & $  3.179  ^{+  0.033 }_{-  0.028}$ &cgs \\ 	
\rhopl \smallskip		& $1.13   ^{+ 0.07}_{-0.08}$	& $  0.52 ^{+ 0.06}_{-0.05}$ &\rhoj 	\\
$a$  \smallskip 		& $  0.0334 ^{+  0.0016}_{-  0.0017}$ 	&$  0.0375 ^{+  0.0007}_{-  0.0008}$ &AU		\\
\teq 			& $ 1480  \pm   10$		&$  1710 \pm   20 $& K	\\
\bottomrule
\end{tabular} 
\end{center}
{\small $^{\dagger}$ BJD$_{\rm TDB}$ --~2\,450\,000.0}\\ 
{\small $^{\ddagger}$ $T_{\rm 14}$: time at transit between 1$^{\rm st}$ and 4$^{\rm th}$ contact}\\ 
\end{table*} 

The planetary properties were determined via an 
 MCMC analysis which simultaneously models the WASP
photometry, the follow-up, high-cadence photometry, together with
CORALIE radial velocity measurements, as described 
in detail by \citet{Cameron2007} and
\citet{Pollacco2008}.
The parameters used by the MCMC analysis are: the epoch
of mid transit $T_{0}$, the orbital period $P$, the fractional change
of flux proportional to the ratio of stellar to planet surface areas
$\Delta F = R_{\rm pl}^2/R_{\star}^2$, the transit duration $T_{14}$,
the impact parameter $b$, the radial velocity semi-amplitude $K_{\rm
  1}$, the stellar host mass \mstar\ calculated in \S\ref{mone},  
the Lagrangian elements \secos\ and \sesin~(where $e$ is the
eccentricity and $\omega$ the longitude of periastron), and the
systemic or centre-of-mass velocity $\gamma$. 
For the treatment of the
stellar limb-darkening, the 4-coefficient model of \citet{Claret2000, Claret2004} was adopted, 
using the corresponding tabulated coefficients for each passband, and 
the stellar spectroscopic properties.  In the case of the WASP photometry,
the $R$-band was used as an approximation. Similarly, the  
$I$ and $V$-bands were used for the TRAPPIST {\it I+z} and blue-blocking filters, respectively.  
The sum of the $\chi^2$ for all input data curves with respect to the models was used as the
goodness-of-fit statistic, and each light curve is weighted such that
the reduced-$\chi^2$ of the best-fit solution is $\sim$1.

An initial MCMC solution with a linear, long-term trend in the radial velocities was explored for both planetary systems
by allowing the systemic velocity to change with time (i.e., $d\gamma/dt \ne 0.$). 
No significant variation of the systemic velocity was found for either planetary system.  Thus, the rest of the analyses are
done assumming no long-term trend in the radial velocities (i.e., $d\gamma/dt = 0.$).   
For each planetary system, four different sets of solutions were
considered: (a) a circular solution assuming that the stellar host is on the main sequence,  
(b) a solution with a free-floating eccentricity and the mass--radius main-sequence constraint,
(c) a circular orbit without the mass--radius constraint, and
(d) an orbit with eccentricity as a free parameter with no main-sequence constraint. 

In the case of both planetary systems, 
when the eccentricity was left as a free paramenter (with and without the main-sequence constraint), 
it converged to a small, non-zero value ($e < 0.02$) for all solutions.  
To assess whether these small eccentricities are real, we performed 
the F-test proposed by \citet[][see their Eq.~27]{Lucy1971}. We find in all instances 
that the resulting eccentricity is spurious. 
The orbit could be truly eccentric to the resulting $e$, but the available data
are unable to differentiate between that and a circular orbit.
Furthermore, the longitude of periastron of these eccentric solutions is close to 90\degree\ or $-90$\degree, 
which could also indicate a spurious eccentricity detection.  
Thus, we adopt circular orbits for both WASP-65b and WASP-75b.

\begin{figure*}
  \centering
   \includegraphics[width=1.0\textwidth]{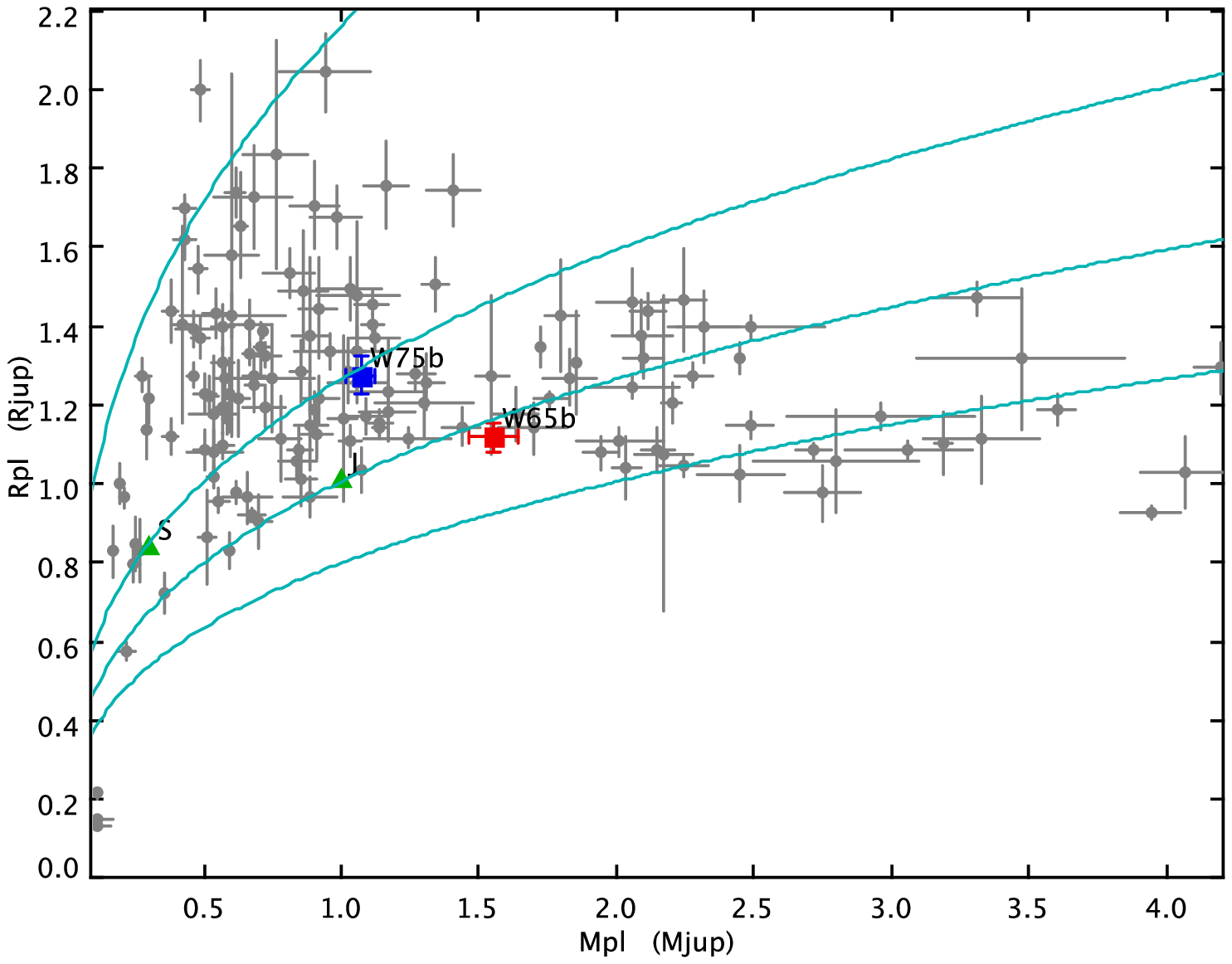}
   \caption{Planet Mass--Radius Diagram. We present two newly discovered planets from the WASP survey for transiting planets, WASP-65b (red-filled square),
and WASP-75b (blue-filled square). 
The grey points represent the known exoplanets in the Saturn and Jupiter mass regimes taken from exoplanet.org (23 May 2013) 
and complemented from the literature.
Saturn and Jupiter are marked in the green-filled triangles, and are initialed.
The continuous (cyan) lines represent equal density traces of 0.1, 0.5, 1.0, and 2.0 \rhoj, from left to right.
WASP-65b lies in between the lower-density giant planets and the higher-density high-mass planets.  
}
\label{massradius}
\end{figure*}

Additionally, we assessed whether either planet host required the assumption of the star being on the main sequence. 
Typically this constraint is needed when the follow-up light curves do not include full transits or do not have the
necessary precision to well determine the transit duration, depth, and the system's impact parameter.
Both WASP-65 and WASP-75 have high-quality follow-up photometry of full transits. 
Comparing both sets of solutions with and without the mass-radius main-sequence assumption, 
the solutions with the main-sequence prior have higher $\chi^2$ values than those without. 
For both planetary systems, both solutions with and without the mass--radius constraint are the same within their 1$-\sigma$ uncertainties;
however the solutions with the main-sequence constraint have larger uncertainties.   
Thus, we adopt the solution of each system that does not impose the main-sequence constraint.

Based on the parameters and considerations described above, 
radius $R$, density $\rho$, and surface gravity $\log g$ of the star (denoted with the subscript $_{\star}$) 
and of the planet (denoted with the subscript $_{\rm pl}$), as well as the mass \mpl\ and the equilibrium temperature of
the planet \teq\ are calculated.   
The planet's equilibrium temperature assumes that it is a black-body ($T_{\rm
  pl,A=0}$) and that the energy is efficiently redistributed from the
planet's day-side to its night-side. 
We also calculate the transit
ingress (and egress) duration $T_{\rm 12}$ ($= T_{\rm 34}$), and the orbital
semi-major axis $a$. 

These calculated properties and their 1--$\sigma$
uncertainties from our MCMC analysis, adopting circular orbits for both planets and
not using the main-sequence constraint, are presented in
Table~\ref{planet_params}.
The corresponding best-fit model radial velocity curves are shown in Fig.~\ref{figrvs}.  
The transit light curve  models are shown in Fig.~\ref{swlcs} against
the WASP observed photometry of WASP-65b (top) and WASP-75b (bottom), in 
Fig.~\ref{w65_lcs} against the follow-up transit light curves of WASP-65b, and 
in Fig.~\ref{w75_lcs} for the WASP-75b follow-up transit light curves.  Each figure
also contains the individual photometric uncertainties and the residuals to the fit.

\section{Discussion}\label{discussion}

We present two newly discovered planets from the WASP survey, WASP-65b and WASP-75b. 

In this paper, we have implemented an estimation of the stellar host mass based
on both theoretical stellar isochrones \citep{Girardi2000,Demarque2004,Pietrinferni2004,VandenBerg2006}, 
and the empirical relationship of \citet{Torres2010} 
as implemented by \citet{Enoch2010} specifically for transiting planets. 
This allows the inclusion of realistic errors in the host mass determination 
emcompassing five independent derivations of the stellar mass
based on the \rhostar\ measured from the transit light curve, and  
the spectroscopically-determined \feh\ and \teff\ including their uncertainties, which are
not always taken into account.  Our analysis includes the propagation of the stellar mass uncertainty in the
planet mass and orbital parameters.  
This is of importance because the planet physical properties are only as accurate as the stellar properties,
and any conclusions directly depend on these derived properties.

WASP-65b has a mass of 1.55 $\pm$ 0.16 \mj, which lies in the mass regime
at which  \citet{Bayliss2013} identify a surprising lack of known hot Jupiters.  
Figure~\ref{massradius} shows this scarcity of 
hot Jupiters at the mass of WASP-65b (red-filled square), and marks
the locus in the mass-radius diagram that separates the lower-density ($\lesssim$1.0\rhoj)
from the higher-density giant planets ($>$1.0\rhoj). 
There are four other known planets with masses consistent with that of WASP-65b within their
1-$\sigma$ uncertainties: WASP-5b, WASP-12b, WASP-50b, and WASP-72b. 
Among these five planets, WASP-65b is the smallest/densest.  It is also 
the one in the orbit with the longest period; though WASP-72b's orbital period ($\sim$2.22d)
is similar to that of WASP-65b.  
Their planet hosts range from 5400 (WASP-50) to 6300~K (WASP-12), and from -0.12 to 0.3 dex in \feh. 
WASP-65 also seems to be the oldest planet host; however, given the uncertainties in the ages,
this is not well constrained. 
It remains unclear whether this paucity of known planets with a mass of $\sim$1.5 \mj\ is the
result of a real physical process that might inhibit the formation of giant planets of this mass, 
a systematic effect in the planetary mass of the known planets because of inaccurate host masses,  
or due to low-number statistics.   The discovery of WASP-65b suggests that we cannot discard any explanation at this time.  
More discoveries of hot Jupiters, in tandem with a careful re-analysis of the known transiting planets,
such as that proposed by \citet{Gomez2013}
will enable the confirmation/rejection of these scenarios. 

The mean density of WASP-65b (1.13 $\pm$ 0.08 \rhoj) is slightly higher than that of Jupiter, and
in fact, WASP-65b is one of the densest planets known in the mass range of  0.1 $<$ \mpl\ $<$ 2.0 \mj.  
WASP-75b is also shown in Fig.~\ref{massradius}, marked by the blue-filled square.  
 With \mpl\ = 1.07 $\pm$ 0.05 \mj, WASP-75b has a {\bf mean} density (0.52 $\pm$ 0.06 \rhoj) similar to that of Saturn. 

Given the measured semi-major axes of their orbits (0.033 $\pm$ 0.002 AU for WASP-65, and 0.0375 $\pm$ 0.0008 AU for WASP-75), 
we compared the theoretical models of \citet{Fortney2007} for planets orbiting a solar-type star at two different orbital separations (0.02 and 0.045 AU).   
We considered the models that predict the largest planet radii: those without a core 
that are composed of only H/He, and those with a 10 \me\ core composed of 50\% rock and 50\% ice and a H/He envelope.  
We find that the radius of WASP-65b (1.11 $\pm$ 0.06 \rj) is not inflated, and is consistent with
all of the predicted radii at 10 Gyr for a 1.5 \mj\ planet in the cases mentioned above. 
 This agreement in the planetary radii could also be considered as evidence of the old age of the system as
suggested by the stellar isochrones ($\sim$9-11 Gyr; see Table~\ref{tracks}), 
given that hot Jupiters {\bf generally decrease in size as they evolve \citep[see e.g., Fig. 5 in][]{Fortney2007}, } 
independent of heating mechanisms \citep[e.g.,][]{Spiegel2013}. 
In the case of WASP-75b, we find the measured radius of 1.27 $\pm$ 0.05 \rj\ to be inflated
by $<$10\% as compared to the coreless models for a 1.0 \mj\ planet with an age of 3.16 Gyr 
orbiting at a distance of 0.02 AU \citep{Fortney2007}.  

\citet{Perna2012} identify 
the equilibrium temperature boundary where the atmospheric heat redistribution starts to be less efficient
to be around $\sim$1500--1700~K.  
WASP-65b (\teq\ $\sim$ 1500~K) and WASP-75b (\teq $\sim$ 1700~K) straddle this boundary, and if their atmospheres
were observable they could provide
insight into the heating/cooling mechanisms of planetary atmospheres.
With current capabilities for transmission spectroscopy, it is not possible to study these atmospheres.
The upper limit of the atmospheric scale height is given for an atmosphere composed of 100\% molecular Hydrogen, and for 
the case of both WASP-65b and WASP-75b, it is only of a few hundreds of kilometres
($\sim$200 and $\sim$450 km, respectively).    
In the case of measuring the emission of the planets, 
 the secondary eclipses could be detectable 
in the K-band and in the Spitzer IRAC 1+2 channels. 
However, it will make 
them interesting targets for future planetary atmospheric studies 
\citep[e.g., JWST and EChO;][]{Gardner2006,Tinetti2012}.

\begin{acknowledgements} 
We would like to thank the anonymous referee for significantly improving this paper. 
  The SuperWASP Consortium consists of astronomers primarily from
  University of Warwick, Queens University Belfast, St Andrews, Keele, Leicester, The Open
  University, Isaac Newton Group La Palma and Instituto de Astrofsica
  de Canarias. The SuperWASP-N camera is hosted by the Issac Newton
  Group on La Palma and WASPSouth is hosted by SAAO. We are grateful
  for their support and assistance. Funding for WASP comes from
  consortium universities and from the UK's Science and Technology
  Facilities Council. Based on observations made with
  the CORALIE Echelle spectrograph
  mounted on the Swiss telescope at ESO La Silla in Chile. 
  TRAPPIST is funded by the Belgian Fund for Scientific Research (Fond National 
de la Recherche Scientifique, FNRS) under the grant FRFC 2.5.594.09.F, with 
the participation of the Swiss National Science Fundation (SNF).
Y. G\'omez Maqueo Chew acknowledges postdoctoral funding support from the Vanderbilt Office of the Provost, through the Vanderbilt Initiative in Data-intensive Astrophysics (VIDA) and through a grant from the Vanderbilt International Office in support of the Vanderbilt-Warwick Exoplanets Collaboration.  
M. Gillon and E. Jehin are FNRS Research Associates.  
L. Delrez is FRIA PhD student of the FNRS.
The research leading to these
  results has received funding from the European Community's Seventh
  Framework Programme (FP7/2007-2013) under grant agreement number
  RG226604 (OPTICON). 
 C. Liebig acknowledges the Qatar Foundation for support from QNRF grant NPRP-09-476-1-078.
A. H.M.J. Triaud is a Swiss National Science Foundation fellow under grant number PBGEP2-145594.  
   \end{acknowledgements}

\bibliographystyle{aa}
\bibliography{w65w75.bib}

\end{document}